\documentclass[aps,reprint,10pt,longbibliography,showpacs,amsmath,amssymb,superscriptaddress]{revtex4-1}

\usepackage{graphicx}% Include figure files
\usepackage{color}
\usepackage{amsmath,amstext,amssymb}
\usepackage{enumitem}
\usepackage{hyperref}
\usepackage[ansinew]{inputenc}
\usepackage{appendix}

\newcommand{\be}{\begin{equation}}
\newcommand{\ee}{\end{equation}}
\newcommand{\bea}{\begin{eqnarray}}
\newcommand{\eea}{\end{eqnarray}}
\newcommand{\bra}[1]{\left\langle#1\right|}
\newcommand{\ket}[1]{\left|#1\right\rangle}

\begin{document}

\title{Entanglement-based dc magnetometry with separated ions}

\author{T.~Ruster}
\author{H.~Kaufmann}
\author{M.~A.~Luda}\thanks{Present address: DEILAP, CITEDEF \& CONICET, J.B. de La Salle 4397, 1603 Villa Martelli, Buenos Aires, Argentina}
\author{V.~Kaushal}
\author{C.~T.~Schmiegelow}\thanks{Present address: Departamento de F\'{i}sica, FCEyN, UBA and IFIBA, Conicet, Pabell\'{o}n 1, Ciudad Universitaria, 1428 Buenos Aires, Argentina}
\author{F.~Schmidt-Kaler}
\author{U.~G.~Poschinger}
\affiliation{Institut f\"ur Physik, Universit\"at Mainz, Staudingerweg 7, 55128 Mainz, Germany}

\begin{abstract}
We demonstrate sensing of inhomogeneous dc magnetic fields by employing entangled trapped ions, which are shuttled in a segmented Paul trap. As \textit{sensor states}, we use Bell states of the type $\ket{\uparrow\downarrow}+\text{e}^{\text{i}\varphi}\ket{\downarrow\uparrow}$ encoded in two $^{40}$Ca$^+$ ions stored at different locations. Due to the linear Zeeman effect, the relative phase $\varphi$ serves to measure the magnetic field difference between the constituent locations, while common-mode fluctuations are rejected. Consecutive measurements on sensor states encoded in the $\text{S}_{1/2}$ ground state and in the $\text{D}_{5/2}$ metastable state are used to separate an ac Zeeman shift from the linear dc Zeeman effect. We measure magnetic field differences over distances of up to $6.2~\text{mm}$, with accuracies of around 300~fT, sensitivities down to $12~\text{pT} / \sqrt{\text{Hz}}$, and spatial resolutions down to $10~\text{nm}$. For optimizing the information gain while maintaining a high dynamic range, we implement an algorithm for Bayesian frequency estimation.
\end{abstract}


\maketitle

\section{Introduction}
Magnetic field sensors are ubiquitous in modern technology and applied and fundamental research. Various sensing technologies are available, covering different parameter regimes in terms of sensitivity, spatial resolution, bandwidth, and other parameters. 
While commonly used devices such as SQUIDs \cite{Jaklevic1964} are already based on quantum effects, recent advances in quantum technology bring genuine \textit{quantum sensors} within the reach of applications. Magnetometers based on single well-isolated atomic systems or ensembles have been demonstrated, where the accumulated phase of a superposition state during an interrogation time $T$ allows for inference of the ambient magnetic field.  Typically, the choice of a sensing platform requires trading sensitivity versus spatial resolution, as ensemble-based systems are more accurate, but also have larger dimensions. Suitable ensemble systems include atomic vapors \cite{WASILEWSKI2010,BUDKER2007}, ultracold atomic gases \cite{KOSCHORRECK2011}, and color centers in diamonds \cite{SIMIN2016,ANGERER2015,WOLF2015}. In contrast, single vacancy centers have been used for high-resolution imaging of magnetic fields \cite{BALASUBRAMANIAN2008}. For single superconducting quantum bits, high sensitivity has been achieved by harnessing strong coupling to magnetic fields \cite{BAL2012}.

A key parameter for quantum magnetic field sensing is the coherence time $T_2$. For longer $T_2$ times, more phase can be accumulated during an experimental cycle, such that sensitivity scales as $1 / \sqrt{T_2}$. Thus, it is crucial to achieve long coherence times while retaining the sensor functionality.

A well-established method for achieving long coherence times is dynamical decoupling \cite{BIERCUK2009}, where the desired signal is spectrally separated from noise. However, this technique is restricted to measurements of alternating magnetic fields. Recently, dynamical decoupling with a single trapped ion has been used to demonstrate magnetometry in the radio-frequency range, attaining a few-pT/$\sqrt{\text{Hz}}$ level of sensitivity \cite{BAUMGART2016,KOTLER2011}. For various applications, it is important to map out the spatial structure of ac magnetic fields \cite{Horsley2015,Warring2013}.

Quantum entanglement can be harnessed to extend sensing capabilities \cite{ROOS2006,Unden2016}. Entangled Greenberger-Horne-Zeilinger or NOON states can in principle yield a sensitivity beyond the standard quantum limit \cite{Huelga1997,Leibfried1476,JONES2009}. However, an increased sensitivity also implies an increased noise-induced decoherence \cite{MONZ2011}. Hence, the beneficial effect of entanglement is generally compromised unless measurement schemes are designed to reject noise in favor of the desired signal. With trapped ions, entangled {\itshape sensor states} of the type $\left(\ket{\uparrow \downarrow } + \text{e}^{\text{i}\varphi} \ket{ \downarrow \uparrow } \right) / \sqrt{2}$ have been employed to measure the magnetic dipole interaction between the constituents' valence electrons \cite{KOTLER2014}.

In this manuscript, we present a sensing scheme for magnetic fields, where entangled ions are moved to different locations $x_1$ and $x_2$ along the trap axis of a segmented linear Paul trap. The dc magnetic field \textit{difference} $\Delta B(x_1,x_2)$ between the ion locations can be inferred from the phase accumulation rate of these sensor states via the linear Zeeman effect
\begin{equation}
 \label{eq:fit}
\Delta \omega(x_1,x_2)_\text{dc} \equiv \dot{\varphi}_\text{dc} = \frac{g \mu_\text{B}}{\hbar} \Delta B(x_1,x_2).
\end{equation}
Since the net magnetic moment of the two constituent ions vanishes, common-mode noise is rejected such that the $T_2$ time exceeds $20~\text{s}$ \cite{HAEFFNER2005}. The long coherence time and the fine-positioning capabilities offered by trapped ions enables us to operate in a parameter regime which could previously not be accessed: We sense \textit{dc} field differences at around 300~fT accuracy and few-pT/$\sqrt{\text{Hz}}$ sensitivity, with spatial resolution down to the few-nanometer range.

In Sec. \ref{sec:expproc}, we describe the procedure for measuring the relative phase $\varphi$ of sensor states, apply it to determine phase accumulation rates $\Delta \omega(x_1,x_2)$ in Sec. \ref{sec:phaseaccmeas}, and discuss the limitations in Sec. \ref{sec:coherencetimes}. An efficient measurement scheme utilizing Bayesian frequency estimation is presented in Sec. \ref{sec:bayes}. Finally, in Sec. \ref{sec:Bdetermination}, we extend our sensing scheme to infer both dc and ac magnetic field differences from the measured phase accumulation rates.

\section{Experimental procedure}
\label{sec:expproc}
We trap two $^{40}\text{Ca}^+$ ions in a segmented linear Paul trap \cite{SCHULZ2008}, featuring 32 control electrode pairs along the trap axis $x$. The distance between the center of neighboring electrodes is $200~\mu \mathrm{m}$.
\begin{figure}[!tb]
	\centering
	\includegraphics[width=0.98\columnwidth]{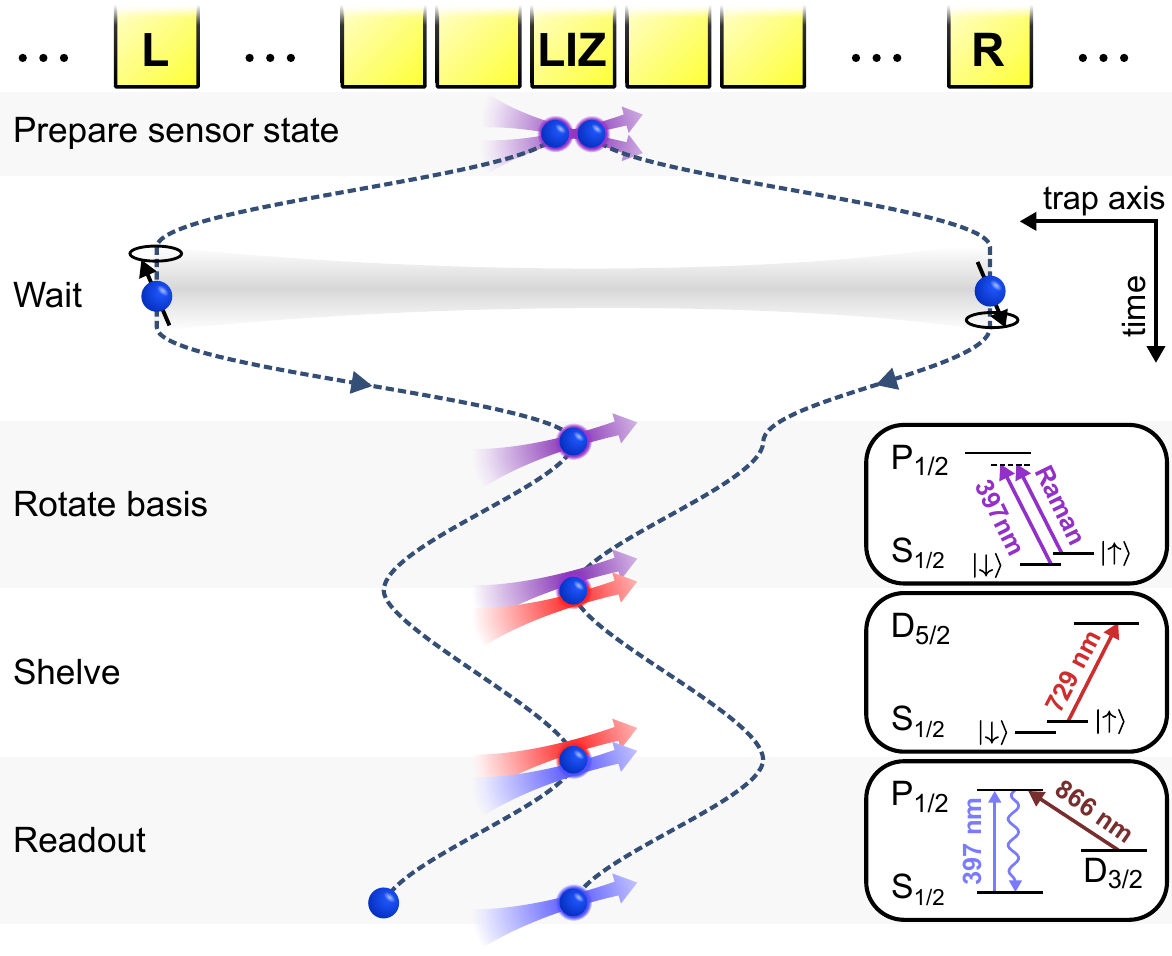}
	\caption{Experimental procedure for measurements of inhomogeneous magnetic fields. After creation of the sensor state at the laser interaction zone (LIZ), the two constituent ions are separated and shuttled to the desired trap segments $L$ and $R$. To measure the accumulated phase during the interrogation time $T$, the ions are individually shuttled to the LIZ to perform basis rotations that allow for state readout via electron shelving and fluorescence detection in either the $\hat{X}_1\hat{X}_2$ or $\hat{X}_1\hat{Y}_2$ basis. For basis rotations, electron shelving, and fluorescence detection, the relevant energy levels are shown.}
	\label{fig:magnetometry_sketch}
\end{figure}
A quantizing magnetic field at an angle of 45$^{\circ}$ to the trap axis is created by $\text{Sm}_2 \text{Co}_{17}$ permanent magnets, splitting the ground state Zeeman sublevels $\ket{\downarrow}\equiv\ket{\text{S}_{1/2},m_j=-\tfrac{1}{2}}$ and $\ket{\uparrow}\equiv\ket{\text{S}_{1/2},m_j=+\tfrac{1}{2}}$ by about $2 \pi \cdot 10.4~\mathrm{MHz}$. The trap setup is shielded from ambient magnetic field fluctuations by a $\mu$-metal magnetic shielding enclosure, yielding a coherence time of about $300~\text{ms}$ \cite{RUSTER2016} in a Ramsey-type experiment.

Laser cooling, coherent spin manipulations and readout \cite{POSCHINGER2009} take place in the laser interaction zone (LIZ) of the trap (Fig. \ref{fig:magnetometry_sketch}). An experimental cycle starts with Doppler laser cooling a two-ion crystal on the $\text{S}_{1/2} \leftrightarrow \text{P}_{1/2}$ \textit{cycling transition} near $397~\mathrm{nm}$. All collective modes of vibration of the ion crystal in radial direction are cooled close to the motional ground state via resolved sideband cooling on the stimulated Raman transition between $\left| \uparrow \right>$ and $\left| \downarrow \right>$. State initialization to $\left| \uparrow \uparrow \right>$ is achieved via frequency-selective pumping utilizing the narrow $\text{S}_{1/2} \leftrightarrow \text{D}_{5/2}$ quadrupole transition near $729~\mathrm{nm}$.

Then, we carry out an entangling gate operation \cite{leibfried2003experimental}, where a spin-dependent optical dipole force is applied to transiently displace the ions in phase space only if the spins are aligned in parallel. To provide spin-motion coupling to the transverse vibrational modes, we direct two orthogonally propagating laser beams to the trap, detuned by $2 \pi \cdot 300~\mathrm{GHz}$ from the cycling transition, such that the difference wave vector is aligned orthogonally to the trap axis. The entangling gate generates a Bell state $\left(\left|\uparrow \uparrow \right> + \left| \downarrow \downarrow\right> \right) / \sqrt 2$. A subsequent $\pi/2$-pulse is applied to acquire the sensor state $\left(\left|\uparrow \downarrow \right> + \left| \downarrow \uparrow\right> \right) / \sqrt 2$, reaching a fidelity of 99.3 (5)\%.

For the subsequent operations, the two-ion crystal is separated \cite{KAUFMANN2014, RUSTER2014}. Ion movement along the trap axis is controlled by applying time-dependent voltages on individual trap electrodes via a fast multichannel arbitrary waveform generator at update rates of up to $2.5~\text{MSamples}/\text{s}$ \cite{WALTHER2012}. After separation, the ions are shuttled to the desired locations $x_1$ and $x_2$ with a maximum distance of $6.2~\text{mm}$. The ions are kept at these locations for an interrogation time $T$. Any inhomogeneity of the magnetic field leads to the accumulation of a phase $\varphi(x_1,x_2,T)$ according to Eq. \ref{eq:fit}, resulting in the state $\left(\left| \uparrow \downarrow \right> + \text{e}^{\text{i}\varphi} \left| \downarrow \uparrow \right> \right) / \sqrt 2$. 

After the interrogation time $T$, both ions are consecutively moved back to the laser interaction zone for spin readout. There, a pair of co-propagating laser beams drives local spin rotations in order to measure the spin along a given basis. Then, population in the state $\left| \uparrow \right>$ is selectively transferred for each ion via laser-driven rapid adiabatic passage to the metastable $\text{D}_{5/2}$ state, followed by conditional detection of resonance fluorescence via a photomultiplier tube while driving the cycling transition (see Fig. \ref{fig:magnetometry_sketch} inset).

Rather than fully reconstructing the quantum state by measuring in 9 different bases, we reduce the number of required measurements by parameterizing the density matrix describing the spin state of the two ions as
\begin{equation}
\hat{\rho}=\frac{1}{2}\begin{pmatrix}
0 & 0 & 0 & 0 \\
0 & 1 & C e^{-i\varphi}& 0 \\
0 & C e^{i\varphi} & 1 & 0 \\
0 & 0 & 0 & 0 
\end{pmatrix}
\label{eq:densitymatrix}
\end{equation}
in the logical basis $\{\ket{\uparrow \uparrow},\ket{\uparrow \downarrow},\ket{\downarrow \uparrow},\ket{\downarrow \downarrow}\}$, with the parity contrast $0\leq C \leq 1$. This form relies on the assumption of balanced populations in the states $\ket{\downarrow\uparrow}$ and $\ket{\uparrow\downarrow}$, which requires that only dephasing and spin rotations about the $Z$ axes take place after state preparation. To infer the phase $\varphi$ and contrast $C$, it is sufficient to measure the parity of the two operators $\hat{X}_1\hat{X}_2$ and $\hat{X}_1\hat{Y}_2$. Thus, upon measuring the operators $\{\hat{X}_1\hat{X}_2,\hat{X}_1\hat{Y}_2\}$ each $\{N,M\}$ times, $\varphi$ and $C$ are determined from the number of events $\{n,m\}$ where the state has been projected to either $\ket{\uparrow\uparrow}$ \textit{or} $\ket{\downarrow\downarrow}$. As described in Appendix \ref{appendix:phase_measurement}, the phase $\varphi$ and contrast $C$ are extracted from the parity results via maximum likelihood estimation. Using numerical simulations, we have confirmed that within the parameter regime and significance level of our measurements, the employed phase estimation method is robust against population imbalance and population leakage to even states.

\label{sec:exp}

\section{Phase accumulation measurements}
\label{sec:phaseaccmeas}
\begin{figure}[tb]
	\centering
	\includegraphics[width=0.98\columnwidth]{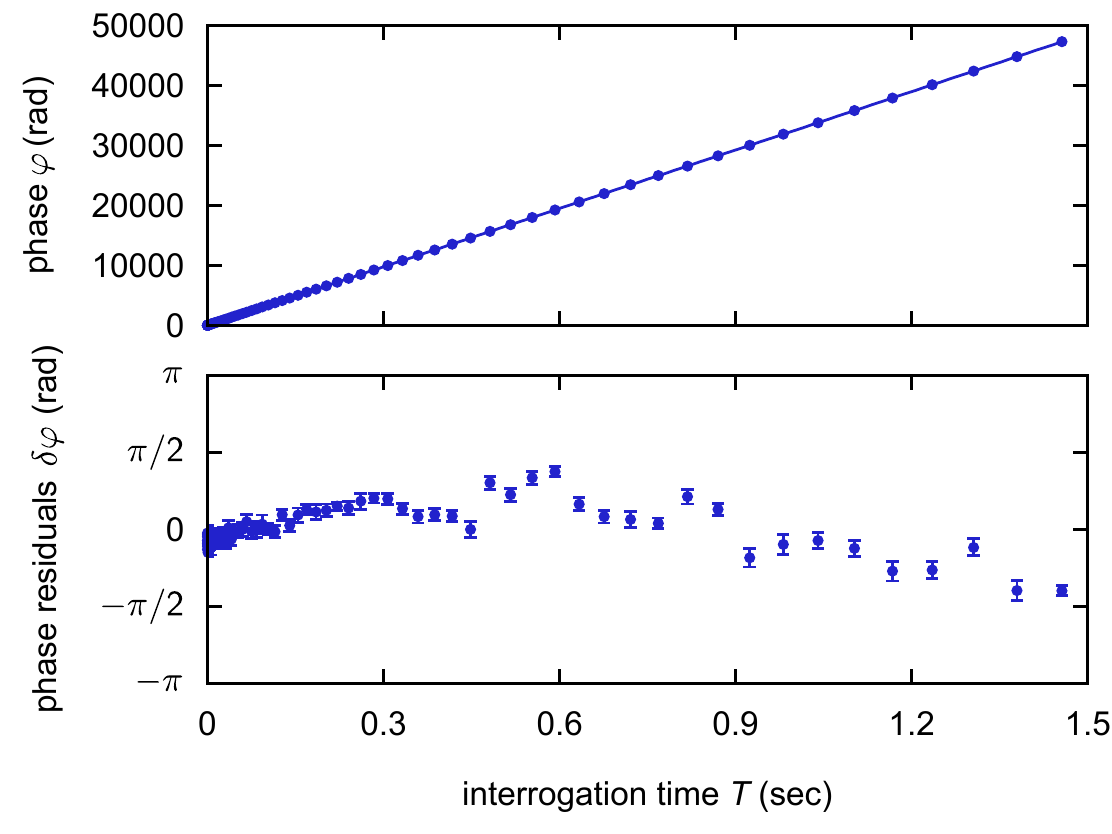}
	\caption{Incremental measurement of the phase accumulation rate $\Delta \omega$ at an ion distance of $d = 6.2~\text{mm}$. A linear fit to measurements of the accumulated phase $\varphi$ at predefined interrogation times (top part), and the fit residuals $\delta \varphi$ for each phase measurement are shown (bottom part). For each point, measurements of both operators have been repeated $50$ times.}
	\label{fig:classical_measurement}
\end{figure}

To determine the phase accumulation rate $\Delta \omega(x_1,x_2)$ of the sensor state with both high sensitivity and high dynamic range, a measurement scheme is required that takes the $2 \pi$-ambiguity of phase measurements into account \cite{waldherr2012high}. In a straightforward incremental approach, we consecutively perform phase measurements at slowly increasing, predefined interrogation times. A linear fit reveals the phase accumulation rate $\Delta \omega(x_1,x_2)$ and a phase offset $\varphi_0(x_1,x_2)$, which arises from ion movement in an inhomogeneous magnetic field. For each phase measurement, the result is incremented or decremented by multiples of $2 \pi$ until it falls within a range of $\pm \pi$ to the previously determined fit function. To check if the phase has been incremented or decremented properly, we verify that the residuals of all points are well below $\pi$. Figure \ref{fig:classical_measurement} shows an example measurement at maximum ion distance $d = 6.2~\text{mm}$ and the residuals $\delta \varphi$ for each point. In this measurement, phases of over $40\,000~\text{rad}$ have been accumulated during interrogation times of up to $T_\text{max} = 1.5~\text{s}$, but the residuals $\left| \delta \varphi \right|$ of all measurement points are well below $\pi$.

The maximum wait time $T_\text{max}$ is ultimately limited by the coherence time $T_\text{coh}$ of the sensor state. For best sensitivity, it is desired to choose $T_\text{max} = T_\text{coh} / 2$ \cite{WOLF2015}. The coherence time is therefore analyzed in the following section.

\section{Coherence times}
\label{sec:coherencetimes}
We characterize the coherence time $T_\text{coh}$ of the sensor state for two settings: (i) The ions are kept in a common harmonic potential well at a distance of about $4.2~\mu\text{m}$, and (ii) for the maximum possible distance of $6.2~\text{mm}$. The coherence time is inferred from measurements of the contrast $C$ for varying interrogation times $T$. For each interrogation time, we repeat the experimental procedure $400$ times for each of the two measurement operators.

\begin{figure}[tb]
	\centering
	\includegraphics[width=0.98\columnwidth]{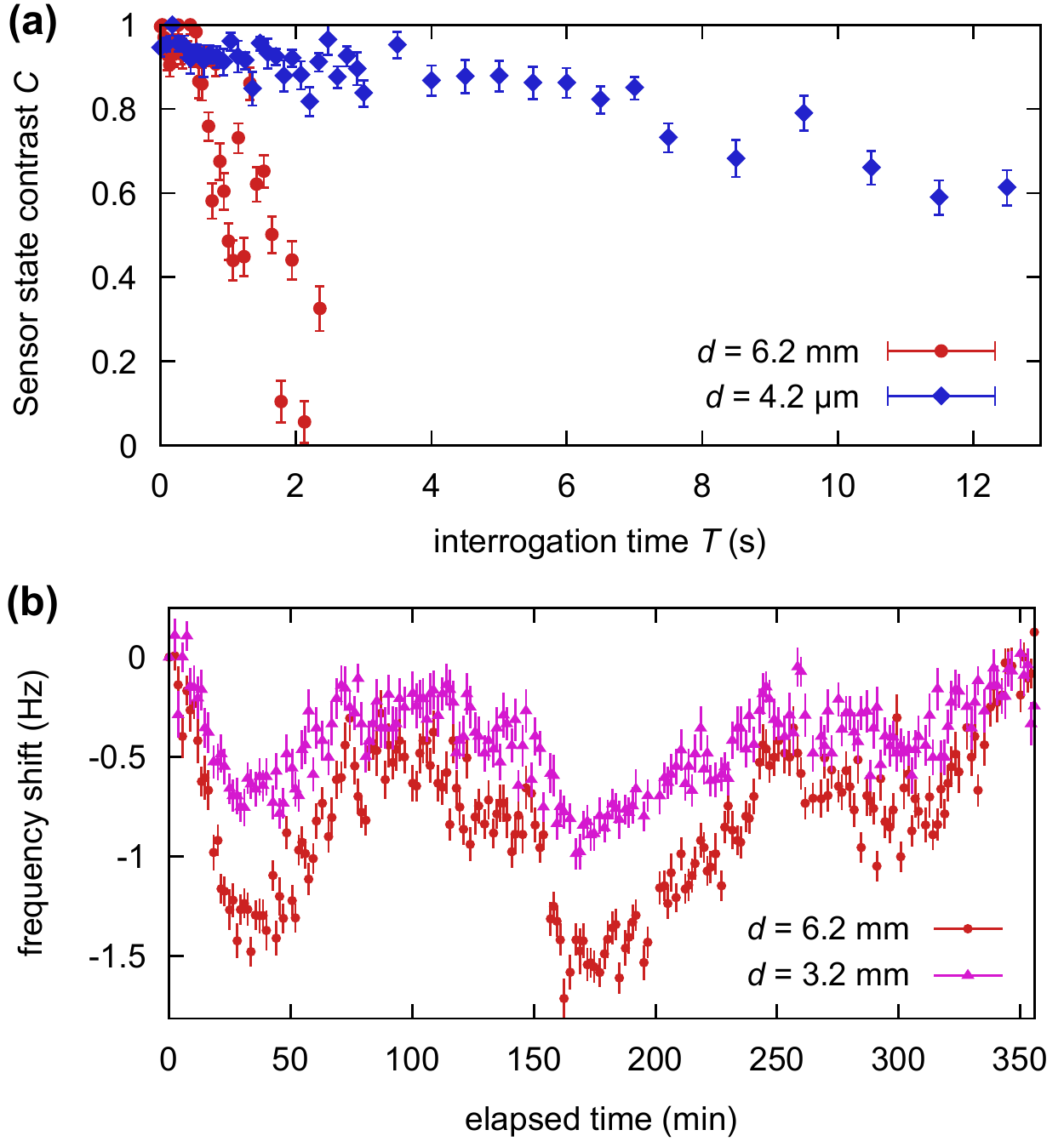}
	\caption{(a) Sensor state contrast $C$ versus interrogation time $T$ at the maximum ion distance of $d=6.2~\text{mm}$ (red dots) and at an ion distance of $d=4.2~\mu\text{m}$ (blue squares). (b) Simultaneous drift of the measured frequency difference for ion distances $d = 6.2~\text{mm}$ (blue circles) and $d = 3.2~\text{mm}$ (purple triangles) over a duration of about 6 hours with an interrogation time of $T=150~\text{ms}$. For $d = 3.2~\text{mm}$, the measured drift is suppressed by a factor of about $1.6$ as compared to the maximum ion distance.}
	\label{fig:coherencetimes}
\end{figure}

For case (i), a coherence time $T_\text{coh} > 12.5~\text{s}$ is observed (Figure \ref{fig:coherencetimes}a). In this regime, residual heating of the radial modes of motion compromises the fidelity of electron shelving, and therefore the spin readout. In a separate measurement, we confirmed that the contrast loss is entirely caused by loss of spin readout fidelity.

\begin{figure*}[tb]
	\centering
	\includegraphics[width=\textwidth]{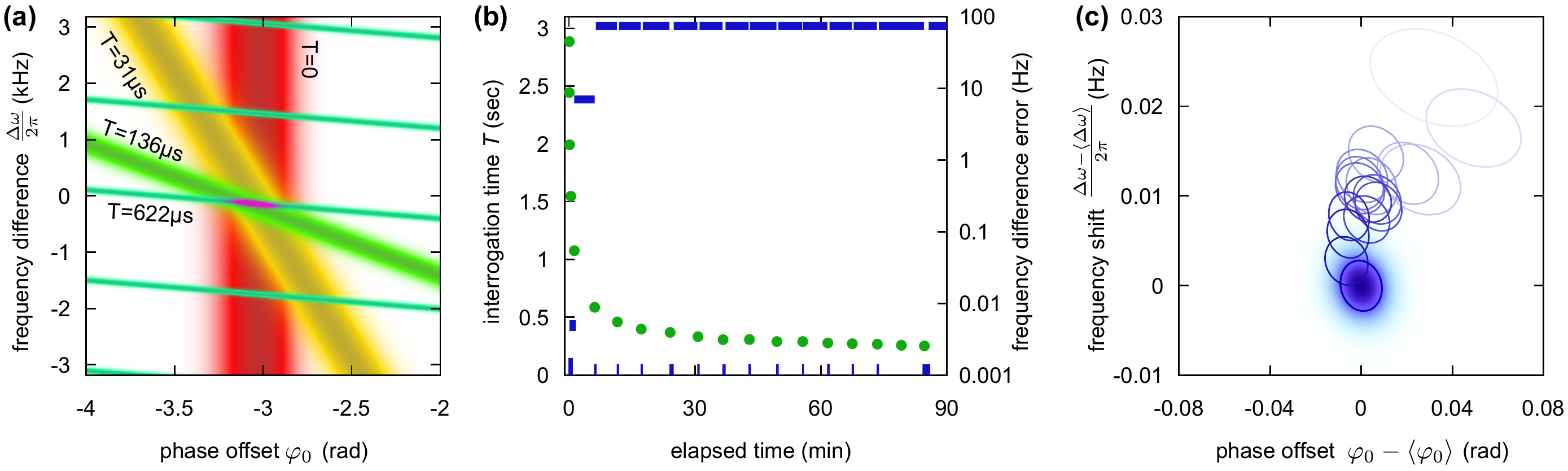}
	\caption{Bayesian evaluation of a measurement at an ion distance of $d=800~\mu\text{m}$. (a) Update functions (colored bars) in parameter space of the first six iterations with $N=M=50$ repetitions of the experimental procedure, and the posterior PDF (purple ellipse) after these iterations. (b) Interrogation times (blue bars) and error of the frequency determination (green points) for each phase measurement versus elapsed measurement time. (c) Final posterior PDF. The posterior PDF of previous phase measurements is visualized by open ellipses, corresponding to the $39.4\%$-credible regions.}
	\label{fig:bayes_measurement}
\end{figure*}

For the maximum possible ion distance, a coherence time in the $1-2~\text{s}$ range has been obtained. The observed coherence loss is presumably caused by a slow drift of the magnetic field inhomogeneity along the trap axis over time, which can be caused by magnetization changes of the permanent magnets, or by movement of the ion trap relative to the magnets. Figure \ref{fig:coherencetimes}b displays these drifts over 6 hours, consecutively measured for two different ion separation distances of $d = 6.2~\text{mm}$ and $d = 3.2~\text{mm}$.

\section{Bayesian frequency estimation}
\label{sec:bayes}
In order to speed up the incremental scheme for determining $\Delta \omega(x_1,x_2)$ described in Sec. \ref{sec:phaseaccmeas}, we dynamically update the interrogation time $T$ for each phase measurement based on previous results. To choose interrogation times that maximize the information gain per measurement cycle, we implement a Bayesian algorithm for frequency estimation \cite{Macieszczak2014,Wiebe2015}.

In Bayesian statistics, for a given phase measurement to be carried out, the combined result of all previous measurements is expressed with the {\itshape prior} probability distribution function (PDF) $p \left( \Delta \omega, \varphi_0\right)$. Initially, we assume a uniformly distributed prior PDF, limited to a reasonable parameter range $\Delta \omega \in \left\{ \Delta \omega_{\text{min}}, \Delta \omega_{\text{max}}\right\}$ and $\varphi_0 \in \left\{ - \pi, \pi\right\}$. After a phase measurement with the outcome $\{n,m\}$, the combined result is described by the {\itshape posterior} PDF
\begin{equation}
	\label{eq:bayesupdate}
	\tilde{p} \left(\Delta \omega, \varphi_0 | n,m; T\right) = \frac{p \left(n,m|\Delta \omega, \varphi_0;T \right) p \left( \Delta \omega, \varphi_0\right)}{ p \left(n,m | T \right)}
\end{equation}
with the {\itshape update function} $p \left(n,m|\Delta \omega, \varphi_0;T \right)$, given by the likelihood function, and the marginal PDF $p \left(n,m | T \right)$ (see Appendix \ref{appendix:bayes_stat}).

The interrogation time $T$ for each phase measurement is calculated such that the expected increase of the Shannon information in the posterior PDF is maximized (see Appendix \ref{appendix:bayes_exp}). With this approach, we observe the automated measurement operate in two distinct measurement regimes: The measurement starts in the {\itshape capture regime}, where $T$ is consecutively increased from $T=0$ to the desired maximum time $T_\text{max}$ in order to unambiguously identify $\Delta\omega$ without any previous information on its value. Then, in the {\itshape tracking regime}, the algorithm alternates $T$ between $T_{\text{max}}$ and $T=0$ for best sensitivity. In order to efficiently track drifts of $\Delta\omega$, we intentionally cause a 'memory loss' by broadening the prior PDF by about 5\% of its width for tracking phase measurements at $T_{\text{max}}$. This facilitates the determination of frequencies which deviate from the previous mean value.

Figure \ref{fig:bayes_measurement} visualizes an example measurement. In Fig. \ref{fig:bayes_measurement}a, the update functions of the first phase measurements in the capture regime are shown. It can be seen that a single phase measurement alone is not sufficient to estimate $\Delta \omega$. However, the combined result of multiple phase measurements yields an approximate Gaussian marginal distribution of $\Delta\omega$, from which the mean value $\left< \Delta \omega \right>$ and the standard error $\Delta \omega_\text{err}$ are inferred. Fig. \ref{fig:bayes_measurement}b depicts the interrogation time $T$ for each experimental cycle and the standard errors of the results, versus the total elapsed time of the measurement. The maximum interrogation time is reached after about 12 minutes, which is about 10 times faster than in the incremental measurement scheme.

In the tracking regime, the precision limit given by the magnetic field inhomogeneity drift rates and the coherence time is reached, and a minimum error of $\Delta \omega_\text{err} = 2 \pi \cdot 2.5~\text{mHz}$ is obtained. Now, the uncertainty after each measurement is no longer reduced, but the parameter estimates are corrected for drifts, see Fig. \ref{fig:bayes_measurement}c.

The shot-noise limited sensitivity describes the minimal frequency change that can be discriminated within unit time:
\begin{equation}
	S_\omega = \Delta \omega_\text{err} \sqrt{T_{\text{tot}}}
\end{equation}
with the standard error of the frequency measurement $\Delta \omega_\text{err}$, that has been achieved during a total experimental time of $T_{\text{tot}}$ \cite{TAYLOR2008}. As the sensitivity depends on the chosen interrogation time, we calculate $S_\omega$ separately for each phase measurement, only taking prior knowledge of the phase offset $\varphi_0$ into account. At an ion distance of $d=800~\mu\text{m}$, a best sensitivity of $S_\omega = 2 \pi \cdot~116~\text{mHz}/\sqrt{\text{Hz}}$ is obtained for an interrogation time of $T_\text{max} = 3.0~\text{s}$. At this interrogation time, we obtained a mean contrast $C$ of about $0.94$ and an average duration of $3.3~\text{s}$ for a single experimental cycle, i.e. about $91\%$ of the measurement time has been utilized for phase accumulation. Thus, we reach $79\%$ of the theoretical standard quantum limit of $1 / \sqrt{T_\text{max}} = 2 \pi \cdot 92~\text{mHz}/\sqrt{\text{Hz}}$. Our results are on par with recent measurements of ac magnetic fields with single ions \cite{BAUMGART2016}, only surpassed by sensors with larger dimensions \cite{WASILEWSKI2010,WOLF2015}.

\begin{figure}[tb]
	\centering
	\includegraphics[width=0.98\columnwidth]{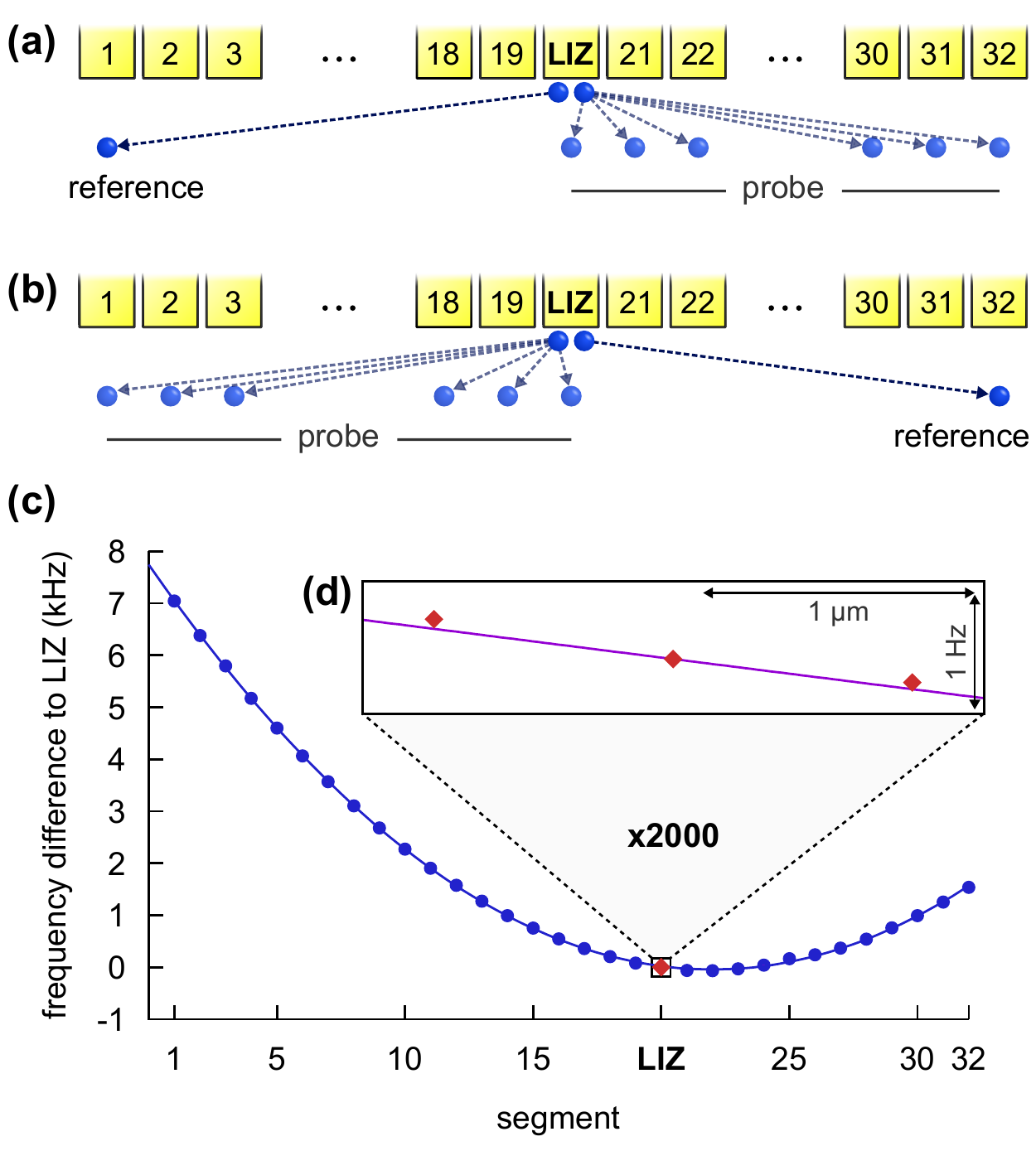}
	\caption{Frequency differences to the laser interaction zone (LIZ). A sensor state is prepared, and the {\itshape probe ion} is moved to an arbitrary desired position. At the same time, the {\itshape reference ion} is either moved to (a) segment 1 or (b) segment 32 to maximize the ion distance. The results for each segment are shown in (c). In (d), high precision measurements close to the LIZ yield standard errors for the ion position and frequency difference of about $10~\text{nm}$ and $2\pi \cdot 10~\text{mHz}$, respectively.}
	\label{fig:magneticfieldmap}
\end{figure}

We utilize our measurement scheme to map the frequency difference $\Delta \omega$ to the laser interaction zone (segment 20) along the trap axis. We perform a frequency measurement for each trap segment, where we move one ion, the {\itshape probe} ion to the desired segment, and the second {\itshape reference} ion to either segment 1 if the probe ion is being moved to segments $20-32$, or to segment 32 if the probe ion is being moved to segments $1-20$ (Fig. \ref{fig:magneticfieldmap}a and \ref{fig:magneticfieldmap}b). This way, the ion distance is sufficiently large such that the trapping potential of the reference ion does not affect the probe ion position at the given level of accurary and vice versa. The results are depicted in Fig. \ref{fig:magneticfieldmap}c. In Fig. \ref{fig:magneticfieldmap}d, additional measurements close to the laser interation zone are shown, allowing to infer frequency gradients with high precision. For reaching spatial resolutions of about $10~\text{nm}$, the probe ion position has been calibrated via an EMCCD camera.

\section{Separation of dc and ac Zeeman shifts}
\label{sec:Bdetermination}
In addition to the \textit{static} Zeeman effect (Eq. \ref{eq:fit}), an additional energy shift is caused by the \textit{ac Zeeman effect} due to oscillating magnetic fields. In our experimental setting, such oscillating fields are generated by the charging/discharging currents of the radiofrequency (rf) electrodes of the Paul trap. In an ideal symmetric trap, the equilibrium positions of the ions are located on the nodal line of the rf field, where also the magnetic fields cancel out. However, residual displacement from the rf node gives rise to a position-dependent frequency shift between the populated magnetic sublevels of a given electronic state \cite{Budker,SUPPLEMENTAL}
\begin{equation}
	\omega^{(\text{ac})}(x) = \Delta m_j \left(g \frac{\mu_\text{B}}{2 \hbar} B_{\text{rf},\bot}(x)\right)^2 \frac{\nu(x)}{\nu(x)^2 - \Omega_\text{rf}^2}. 
\label{eq:acZeemanS}
\end{equation}
Here, $x$ is the ion position along the trap axis, $B_{\text{rf},\bot}(x)$ is the component of the oscillating magnetic field perpendicular to the static quantizing magnetic field, $\Omega_\text{rf} = 2\pi \cdot 33~\text{MHz}$ is the trap drive frequency, and $\nu(x)$ denotes the total (angular) frequency splitting between neighboring ($\vert\Delta m_j\vert=1$) Zeeman sublevels.

\begin{figure}[tb]
	\centering
	\includegraphics[width=0.98\columnwidth]{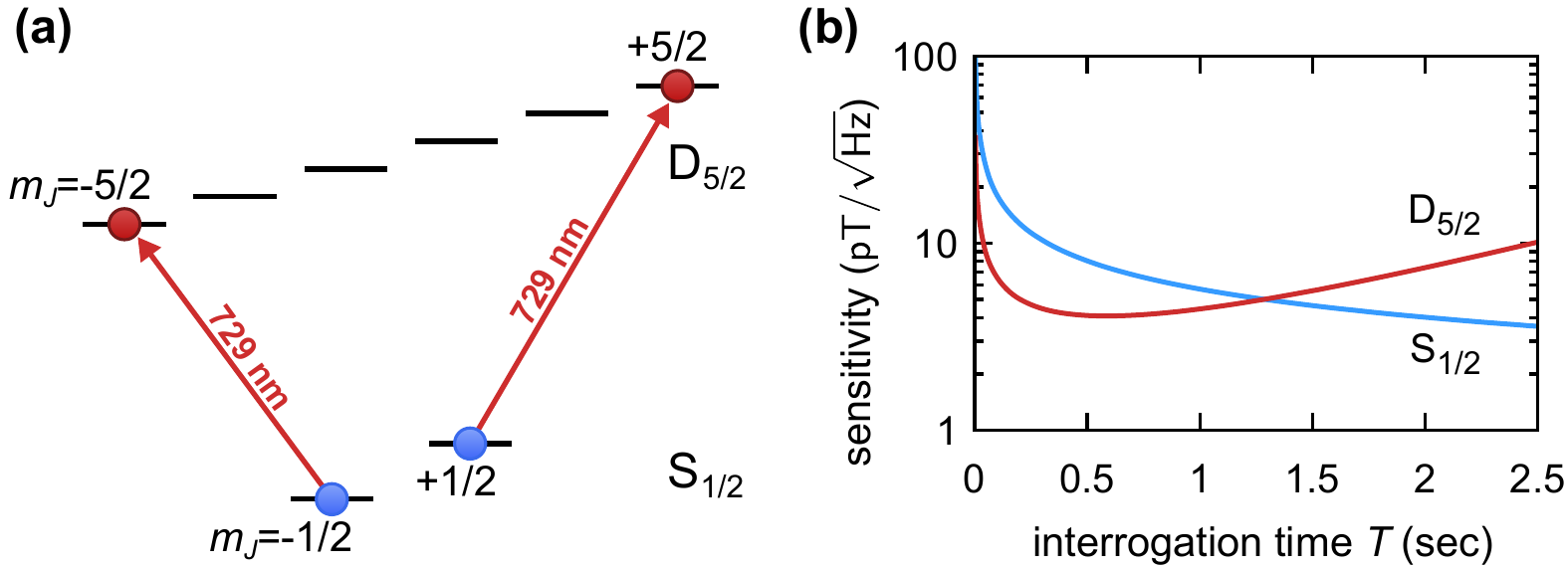}
	\caption{(a) Relevant transitions to create the $\text{D}_{5/2}$ sensor state. (b) Theoretical shot-noise limited sensitivity for magnetic field measurements versus interrogation time $T$ for both $\text{S}_{1/2}$ and $\text{D}_{5/2}$ sensor states.}
	\label{fig:shelved_bell}
\end{figure}

For sensor states encoded in different electronic state manifolds, the respective Land\'e factors lead to different contributions to the total phase accumulation rates from dc and ac fields. Hence, by encoding entangled sensor states within \textit{different} electronic states of $^{40}\text{Ca}^+$, our sensing scheme is extended to distinguish between ac and dc magnetic fields. We utilize the $m_j =\pm 5/2$ sublevels of the metastable $\text{D}_{5/2}$ state in addition to the $\text{S}_{1/2}$ ground state for frequency-difference measurements. We prepare the sensor state $\ket{+5/2,-5/2} + \ket{-5/2,+5/2}$ by first preparing the state $\ket{\uparrow \downarrow } + \ket{ \downarrow \uparrow }$, and then transferring the populations of both ions to the respective sublevels of the $\text{D}_{5/2}$ metastable state, i.e. $\ket{\uparrow}\rightarrow\ket{+5/2}$ and $\ket{\downarrow}\rightarrow\ket{-5/2}$ (Fig. \ref{fig:shelved_bell}a). The population transfer is carried out via composite inversion laser pulses near $729~\text{nm}$ \cite{FREEMAN1980}. Considering the Land\'{e} factors of both states $g_{\text{S}} = 2.00225664(9)$ \cite{TOMMASEO2003} and $g_{\text{D}} = 1.2003340(3)$ \cite{CHWALLA2009}, the $\text{D}_{5/2}$ sensor state features phase accumulation rates which are increased by a factor of $3$. However, spontaneous decay at a rate of $1/\tau$ \textit{per ion} has to be taken into account, with a time constant of about $\tau = 1.17~\text{s}$ \cite{KREUTER2005}. We employ an additional fluorescence detection step before state readout to reject measurements where at least one ion has decayed from the $\text{D}_{5/2}$ state. Beyond wait times of $\tau/2$, this postselection reduces the sensitivity of the measurement (Fig. \ref{fig:shelved_bell}b). 

As the measured differential phase accumulation rates $\Delta\omega_\text{S}(x_1,x_2)$ and $\Delta\omega_\text{D}(x_1,x_2)$ of the respective $\text{S}_{1/2}$ and $\text{D}_{5/2}$ sensor states are affected differently by the static dc Zeeman effect and the ac Zeeman shift, we can infer the magnetic field difference via (see Supplemental material \cite{SUPPLEMENTAL})
\begin{equation}
\Delta B(x_1,x_2) = \frac{\hbar}{\mu_\text{B}} \frac{\Delta \omega_{\text{D}}(x_1,x_2) - \chi \Delta \omega_{\text{S}}(x_1,x_2)}{5 g_\text{D} - \chi g_\text{S}}. \label{eq:deltaB} 
\end{equation}
The differential ac Zeeman shift between the constituent ions of the $\text{S}_{1/2}$ sensor state is given by
\begin{equation}
\Delta \omega^{(\text{ac})}_\text{S}(x_1,x_2) = \Delta\omega_\text{S}-g_\text{S}\frac{\Delta\omega_\text{D}-\chi \Delta\omega_\text{S}}{5 g_\text{D}-\chi g_\text{S}}.\label{eq:deltaOmegaAC}
\end{equation}
Here, $\chi=\Delta \omega^{(\text{ac})}_\text{D}(x_1,x_2)/\Delta \omega^{(\text{ac})}_\text{S}(x_1,x_2)$ denotes the ratio of the differential ac Zeeman shifts pertaining to the $\text{D}_{5/2}$ and $\text{S}_{1/2}$ sensor states. Under the approximation that the magnetic field inhomogeneity is small compared to the absolute magnetic field, i.e. the energy splittings $\nu_\text{S}(x)$ and $\nu_\text{D}(x)$ of the respective electronic states are constant along the trap axis, $\chi$ is calculated via
\begin{equation}
\chi \approx 5 \left(\frac{g_{\text{D}}}{g_{\text{S}}}\right)^2 \frac{\nu_\text{D}}{\nu_\text{S}} \cdot \frac{\nu_\text{S}^2 -  \Omega_\text{rf}^2}{\nu_\text{D}^2 - \Omega_\text{rf}^2}. \label{eq:chiApprox}
\end{equation}
This approximation is well fulfilled in our experimental setup.

\begin{figure}[tb]
	\centering
	\includegraphics[width=0.98\columnwidth]{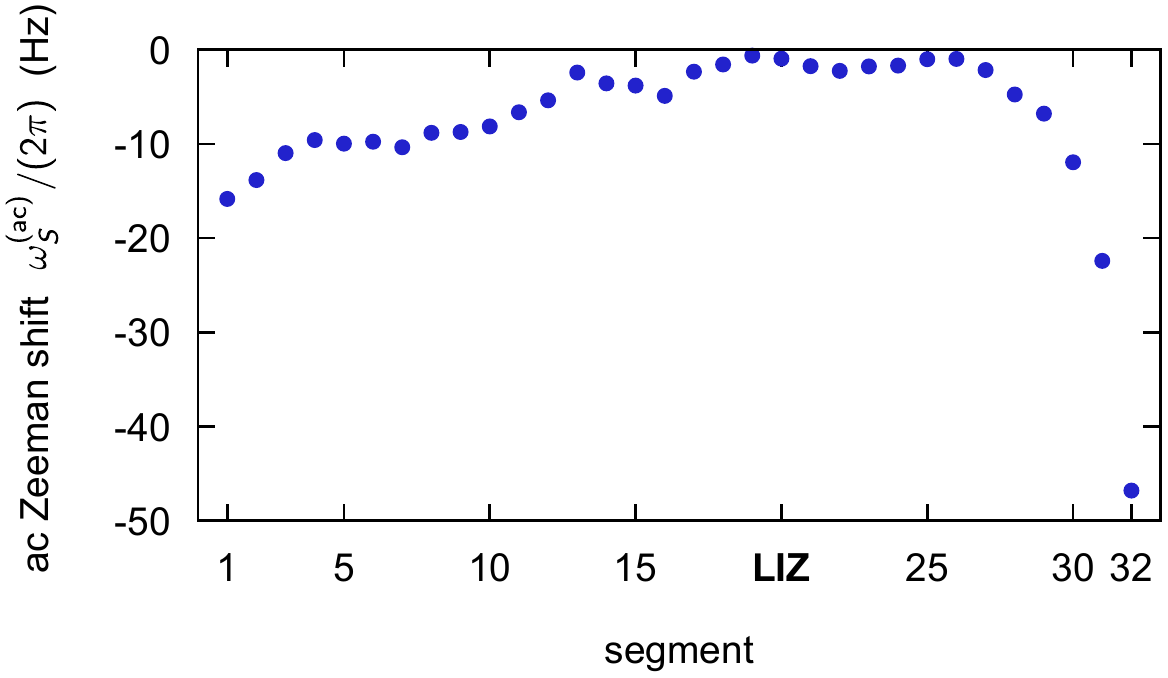}
	\caption{Frequency shift of the $\text{S}_{1/2}$ sensor state due to the ac Zeeman effect along the trap axis. Within the error bars of about $2 \pi \cdot 0.2~\text{Hz}$, the ac Zeeman shift is similar for both $\text{S}_{1/2}$ and $\text{D}_{5/2}$ sensor states.}
	\label{fig:acZeeman}
\end{figure}

Experimentally, we measure the phase accumulation rates $\Delta\omega_\text{S}(x_1,x_2)$ and $\Delta\omega_\text{D}(x_1,x_2)$ by performing alternating experimental cycles on the $S_{1/2}$ and $D_{5/2}$ sensor states. The respective interrogation times $T_{\text{S}}$ and $T_{\text{D}}$ are individually determined by the Bayesian algorithm. Additional measurements on a single ion at the laser interaction zone are employed to determine the transition frequencies $\nu_\text{S}(x_\text{LIZ})$, $\nu_\text{D}(x_\text{LIZ})$, and the absolute ac Zeeman shift $\omega^{(\text{ac})}_\text{S}(x_\text{LIZ})$ (see Supplemental material \cite{SUPPLEMENTAL}). The transition frequencies $\nu_\text{S}(x_\text{LIZ})$ and $\nu_\text{D}(x_\text{LIZ})$ are plugged into the ac Zeeman ratio $\chi$ (Eq. \ref{eq:chiApprox}), which is used to infer $\Delta B(x_1,x_2)$ (Eq. \ref{eq:deltaB}) and $\Delta\omega_S^{(ac)}(x_1,x_2)$ (Eq. \ref{eq:deltaOmegaAC}).

Figure \ref{fig:acZeeman} depicts the absolute ac Zeeman shift along the trap axis. At the laser interaction zone, an ac Zeeman shift of $\omega^{(\text{ac})}_\text{S}(x_\text{LIZ}) = - 2 \pi \cdot 0.93 (12)~\text{Hz}$ is revealed. For remote segments, the magnitude of the frequency shift increases by up to $2 \pi \cdot 50~\text{Hz}$. This behavior is presumably caused by a displacement of the ions' equilibrium positions from to the nodal line of the rf field, which are minimized only at the laser interaction zone to compensate excess micromotion. For all ion positions, standard errors of about $\omega^{(\text{ac})}_\text{S,err} = 2 \pi \cdot 0.2~\text{Hz}$ are reached. Compared to recent measurements of the ac Zeeman shift arising from microwave fields \cite{Warring2013}, this is a 1000-fold improvement in accuracy. Thus, our measurement technique may be used to improve the fidelity of microwave-driven quantum gates, where precise mapping of the ac Zeeman shift is important.

For dc magnetic field differences, we attain sensitivities down to $S_B = 12~\text{pT} / \sqrt{\text{Hz}}$ at interrogation times of $T_{\text{S}} = 1.50~\text{s}$ and $T_{\text{D}}= 0.48~\text{s}$. Accuracies as good as $\Delta B^{(\text{stat})}_\text{err} = 310~\text{fT}$ are reached at an ion distance of $d=800~\mu\text{m}$.

\section{Conclusion and Outlook}
We have demonstrated a novel magnetometry scheme harnessing entangled ions, which are freely positioned in a segmented Paul trap. The long coherence time of the entangled states enable precise measurement of dc magnetic field differences. Our measurement scheme additionally characterizes the position-dependent ac Zeeman effect due to the rf trap drive in Paul traps, which is a hard-to-characterize source of errors for precision measurements in frequency standards. For recent optical clocks, the ac Zeeman shift contributes to the fractional error in the $10^{-20}$-$10^{-17}$ range \cite{Herschbach2012, Chou2010, Itano2007}.

Precise knowledge of the magnetic field along the trap axis is essential for a shuttling-based approach towards scalable quantum information experiments in Paul traps. In this approach, quantum algorithms are carried out with multiple ions residing at different trap segments, where different phases are accumulated. These phases have to be taken into account within the computational sequences \cite{Kaufmann2016}.

With the presented measurement technique, it will be feasible to characterize the magnetic field of objects close to or inside the trap volume, such as neutral atoms trapped by optical dipole forces, or additional trapped ions \cite{SchmidtKaler2012}. The current limitation of our magnetometry scheme is given by magnetic field drifts, which can be mitigated by using a spatially homogeneous quantizing magnetic field and temperature-stabilized permanent magnets.

%%%%%%%%%%%%%%%%%%%%%%%%%%%%%%%%%%%%%%%%%%%%%%%%%%%%%%%%%%%%%%%%%%%%
%\section{Acknowledgements}
%%%%%%%%%%%%%%%%%%%%%%%%%%%%%%%%%%%%%%%%%%%%%%%%%%%%%%%%%%%%%%%%%%%%
\begin{acknowledgments}
We acknowledge former contributions of Claudia Warschburger and helpful discussions with Georg Jacob, Arne Wickenbrock and Lykourgos Bougas. We further acknowledge support from the DFG through the DIP program (Grant No. SCHM 1049/7-1) and the Bundesministerium f\"ur Bildung und Forschung via IKT 2020 (Q.com). The research is based upon work supported by the Office of the Director of National Intelligence (ODNI), Intelligence Advanced Research Projects Activity (IARPA), via the U.S. Army Research Office grant W911NF-16-1-0070. The views and conclusions contained herein are those of the authors and should not be interpreted as necessarily representing the official policies or endorsements, either expressed or implied, of the ODNI, IARPA, or the U.S. Government. The U.S. Government is authorized to reproduce and distribute reprints for Governmental purposes notwithstanding any copyright annotation thereon. Any opinions, findings, and conclusions or recommendations expressed in this material are those of the author(s) and do not necessarily reflect the view of the U.S. Army Research Office.
\end{acknowledgments}

\appendix
\section{Contrast and phase estimation}
\label{appendix:phase_measurement}
As explained in Sec. \ref{sec:exp}, the outcome of a measurement at a chosen interrogation time $T$ is described by the parity of the projected state. Assuming the state $\hat \rho$ (Eq. \ref{eq:densitymatrix}), the probabilities to detect even parity for the $\hat{X}_1\hat{X}_2$ and $\hat{X}_1\hat{Y}_2$ measurements are given by
\begin{align}
\begin{split}
p^{(E)}_{XX} &= \text{Tr}\left(\hat{R}_{Y,1}\left(\tfrac{\pi}{2}\right) \hat{R}_{Y,2}\left(\tfrac{\pi}{2}\right)\hat{\rho}\hat{R}_{Y,1}^{\dagger}\left(\tfrac{\pi}{2}\right) \hat{R}_{Y,2}^{\dagger}\left(\tfrac{\pi}{2}\right)\hat{P}_E\right) \\
&= \frac{1}{2}\left( 1 - C \cos\left(\varphi\right)\right) \\
p^{(E)}_{XY} &= \text{Tr}\left(\hat{R}_{Y,1}\left(\tfrac{\pi}{2}\right) \hat{R}_{X,2}\left(-\tfrac{\pi}{2}\right)\hat{\rho}\hat{R}_{Y,1}^{\dagger}\left(\tfrac{\pi}{2}\right) \hat{R}_{X,2}^{\dagger}\left(-\tfrac{\pi}{2}\right)\hat{P}_E\right) \\
&= \frac{1}{2}\left( 1 + C \sin\left(\varphi\right)\right),
\end{split}
\end{align}
where $\hat{P}_E=\ket{\uparrow\uparrow}\bra{\uparrow\uparrow}+\ket{\downarrow\downarrow}\bra{\downarrow\downarrow}$ is the projector onto the subspace of even spin configurations, and $\hat{R}_{X/Y,i}(\theta)$ represent single qubit rotations by angle $\theta$ on ion $i=1,2$ prior to readout. The readout is dichotomic in terms of even/odd spin configurations, and the measurements are independent. Probing operators $\{\hat{X}_1\hat{X}_2,\hat{X}_1\hat{Y}_2\}$ each $\{N,M\}$ times, the probability to observe $\{n,m\}$ even spin configurations for given parameters $\left(\varphi,C\right)$ is given by binomial statistics:
\begin{align}
\begin{split}
\label{eq:binomprobs}
	p_{XX}\left(n|\varphi,C\right) &= \begin{pmatrix}N \\ n\end{pmatrix} {p_{XX}^{(E)}}^n \left(1 - p_{XX}^{(E)}\right)^{N-n} \\
	p_{XY}\left(m|\varphi,C\right) &= \begin{pmatrix}M \\ m\end{pmatrix} {p_{XY}^{(E)}}^m \left(1 - p_{XY}^{(E)}\right)^{M-m}
\end{split}
\end{align}
For a measurement result $\{n,m\}$, the phase $\left<\varphi\right>$ and contrast $\left<C\right>$ are obtained by maximizing the likelihood function
\begin{equation}
L \left(\varphi, C \right) = L \left(n,m;\varphi, C\right)=p_{XX}\left(n|\varphi,C\right)p_{XY}\left(m|\varphi,C\right)
\label{eq:likelihood}
\end{equation}
with regards to $\varphi$ and $C$. If the sample sizes $N$ and $M$ are large, the likelihood ratio 
\begin{equation}
R(\varphi, C) = 2 \log \left(\frac{L \left(\varphi, C \right)}{L \left(\left<\varphi\right>,\left< C\right>\right)} \right)
\end{equation}
is approximately $\chi^2$-distributed, such that $68.3\%$-confidence intervals can be obtained via $R(\varphi, \left<C\right>) \leq 1$ for $\varphi$ and $R(\left<\varphi\right>, C) \leq 1$ for $C$.
%\begin{equation}
%\langle\phi\rangle=\mathcal{N}^{-1}\int_0^1 dc\ c \int_0^{2\pi}d\phi\ \phi\ \phi\ p_{XX}\left(n|\phi,c\right)p_{XY}\left(m|\phi,c\right)
%\end{equation}
%with the normalization factor $\mathcal{N}$. Here, care has to be taken to avoid spurious results if the true value of $\phi$ lies close to the jump between $0$ and $2\pi$. For obtaining a confidence interval, we utilize the Fisher information
%\begin{eqnarray}
%\mathcal{I}_{XX}&=&\left\langle \frac{\partial^2}{\partial\phi^2}\log p_{XX}\left(n|\phi,c\right) \right\rangle \nonumber \\
%							&=&\sum_{n=0}^N \frac{\partial^2}{\partial\phi^2}\log p_{XX}\left(n|\phi,c\right) \nonumber \\
%\mathcal{I}_{XY}&=&\left\langle \frac{\partial^2}{\partial\phi^2}\log p_{XY}\left(m|\phi,c\right) \right\rangle \nonumber \\
%							&=&\sum_{m=0}^M \frac{\partial^2}{\partial\phi^2}\log p_{XY}\left(m|\phi,c\right) \nonumber \\
%\end{eqnarray}
%to invoke the Cram{\'e}r-Rao bound
%\begin{equation}
%\text{Var}(\phi)\geq 1/(\mathcal{I}_{XX}+\mathcal{I}_{XY})
%\end{equation}

\section{Bayesian statistics}
\label{appendix:bayes_stat}
In Bayesian statistics, the result after each phase measurement is described by the posterior PDF $\tilde{p} \left(\Delta \omega, \varphi_0 | n,m; T\right)$ (Eq. \ref{eq:bayesupdate}). The update function is given by
\begin{equation}
	p \left(n,m|\Delta \omega, \varphi_0;T \right) = \int_{0}^{1} L \left(n,m;\varphi(\Delta \omega, \varphi_0; T), C\right)  \ \text{d} C
\end{equation}
For each parameter set $\left( \Delta \omega, \varphi_0\right)$, the accumulated phase after the interrogation time $T$ is given by 
\begin{equation}
 \varphi(\Delta \omega, \varphi_0; T) = \Delta \omega \cdot T + \varphi_0
\end{equation}
Due to the phase periodicity, the update function features a $2 \pi /T$ periodicity in $\Delta \omega$. However, if the width of the prior PDF is smaller than the periodicity of the update function, the periodicity is not inherited by the posterior PDF.
After at least two phase measurements at different interrogation times, the posterior PDF is well described by a two-dimensional normal distribution. To obtain estimates for $\Delta \omega$ and $\varphi_0$, we calculate expectation values from the marginalized PDF:
\begin{align}
	\langle\Delta \omega\rangle & = \int \int \Delta\omega \cdot \tilde{p} \left(\Delta \omega, \varphi_0 | n,m; T\right) \text{d}\Delta \omega \,\text{d} \varphi_0 \\
	\langle\varphi_0\rangle & = \int \int \varphi_0 \cdot \tilde{p} \left(\Delta \omega, \varphi_0 | n,m; T\right) \text{d}\Delta \omega \,\text{d}\varphi_0
\end{align}
Standard errors are obtained in a similar way by calculating the corresponding standard deviations.

\section{Bayesian experimental design}
\label{appendix:bayes_exp}
To calculate the optimal interrogation time $T$ for the next measurement to be performed, we employ Bayes' rule to calculate the posterior PDF for a \textit{hypothetical} measurement result $\{n,m\}$ at interrogation time $T$ with fixed contrast $C$:
\begin{equation}
	\tilde{p} \left(\Delta \omega, \varphi_0 | n,m; C,T\right) = \frac{p \left(n,m|\Delta \omega, \varphi_0;C,T \right) p \left( \Delta \omega, \varphi_0\right)}{ p \left(n,m | C,T \right)}
\end{equation}
with the marginal PDF
\begin{align}
\begin{split}
 p \left(n,m | C,T \right) &= \int \int p \left(n,m|\Delta \omega, \varphi_0;C,T \right) \\
	& \qquad \qquad \times\ p \left( \Delta \omega, \varphi_0\right) \text{d}\Delta \omega\,\text{d}\varphi_0.
\end{split}
\end{align}
Here, it is sufficient to consider $N=M=1$ to save computational effort. Because we are interested in minimizing the error in $\Delta \omega$, we marginalize
\begin{equation}
	\tilde{p} \left(\Delta \omega \right) := \int \tilde{p} \left(\Delta \omega, \varphi_0 | n,m; C,T\right) \text{d} \varphi_0.
\end{equation}
%\subsection{Utility}
{\itshape Utility} is defined as the expected gain in Shannon information of the posterior with respect to the prior after a hypothetical measurement
\begin{equation}
U \left(n,m; T \right) = \int \tilde{p} \left(\Delta \omega \right) \log \tilde{p} \left(\Delta \omega \right) \text{d} \Delta \omega  - U_0,
\end{equation}
with the Shannon information of the marginalized prior PDF
%note U_0 is NOT irrelevant here. We need it to correctly apply the penalty factor w(t)
\begin{equation}
U_0 = \int  p \left(\Delta \omega \right) \log p \left(\Delta \omega \right) \text{d}\Delta \omega.
\end{equation}
Then, we average the utility function over all possible measurement results, weighted with the respective marginal probability:
\begin{equation}
U(T) = \sum_{n=0}^N \sum_{m=0}^M w(T) \cdot U\left(n,m;T\right)~p(n,m|C,T)
\end{equation}
Here, a penalty factor $w(T) = D(0) / D(T)$ takes the increased measurement duration for longer interrogation times into account, where $D(T)$ is the duration of a single experimental run with a given $T$.
The ideal interrogation time for an upcoming measurement is $T_0 = \max_T U(T)$, i.e. $T_0$ maximizes the expected gain in Shannon information. Via the known results from the prior PDF $\left<\Delta \omega \right>$ and $\left<\varphi_0\right>$, we add a phase offset to the second $X$ or $Y$ analysis pulse, such that the measured phase is always close to $\pi / 4$. Near $\pi / 4$, the error bar of a single phase measurement is minimized (at the expense of an increased contrast uncertainty). 

\bibliography{longDistance}

\end{document}